\documentclass[letterpaper, 10 pt, conference]{ieeeconf}  

\IEEEoverridecommandlockouts                              

\usepackage{cite}
\usepackage{amsmath,amssymb,amsfonts}
\usepackage{algorithm}
\usepackage{graphicx}
\usepackage{textcomp}
\usepackage{graphics} 
\usepackage{epsfig} 
\usepackage{mathptmx} 
\usepackage{times} 
\usepackage{mathrsfs}
\usepackage{mathtools}
\usepackage{stfloats}
\usepackage[noend]{algpseudocode}
\usepackage{multirow}
\usepackage{subcaption}
\usepackage{url}
\usepackage{xcolor}
\usepackage[bookmarks=false]{hyperref}

\graphicspath{{figures/}}

\title{Beacon-based Distributed Structure Formation in Multi-agent Systems}

\author{Tamzidul Mina$^{1}$, Wonse Jo$^{2}$, Shyam S. Kannan$^{2}$, and Byung-Cheol Min$^{2}$
\thanks{This paper is based on research supported by the National Science Foundation (NSF) under Grant No. IIS-1846221. Any opinions, findings, and conclusions or recommendations expressed in this material are those of the authors and do not necessarily reflect the views of the National Science Foundation.}
\thanks{$^{1}$ Author currently affiliated with Sandia National Laboratories, Albuquerque, NM 87123. This research in its entirety was conducted while previously affiliated with the SMART Lab, Department of Computer and Information Technology, Purdue University, and the School of Mechanical Engineering, Purdue University, West Lafayette, IN 47907, USA. 
        {\tt\small tmina@sandia.gov}}%
\thanks{$^{2}$ SMART Lab, Department of Computer and Information Technology, Purdue University,
        West Lafayette, IN 47907, USA
        {\tt\small \{jow, kannan9, minb\}@purdue.edu}}%
}

\begin{document}

\maketitle
\thispagestyle{empty}
\pagestyle{empty}

\begin{abstract}
Autonomous shape and structure formation is an important problem in the domain of large-scale multi-agent systems. 
In this paper, we propose a 3D structure representation method and a distributed structure formation strategy where settled agents guide free moving agents to a prescribed location to settle in the structure. 
Agents at the structure formation frontier looking for neighbors to settle act as beacons, generating a surface gradient throughout the formed structure propagated by settled agents.
Free-moving agents follow the surface gradient along the formed structure surface to the formation frontier, where they eventually reach the closest beacon and settle to continue the structure formation following a local bidding process. 
Agent behavior is governed by a finite state machine implementation, along with potential field-based motion control laws. We also discuss appropriate rules for recovering from stagnation points.
Simulation experiments are presented to show planar and 3D structure formations with continuous and discontinuous boundary/surfaces, which validate the proposed strategy, followed by a scalability analysis.   

\end{abstract}

\section{Introduction}\label{sec:introduction}
Autonomous structure formation by large-scale multi-agent systems has enormous potential in applications ranging from dangerous construction work such as building habitats for humans on other planets, building make-shift dams for water diversion/containment during flooding in search and rescue operations, to developing programmable matter for manufacturing in the future. 
While multi-agent self-organization has been a popular research area over the years, a feasible implementation strategy remains unrealized due to the inherent challenges involved in the process.

Mora \textit{et al.} \cite{alonso2011multi} identified two fundamental problems in the multi-agent structure formation task: 1) assigning an arbitrary number of agents to goal positions that define the structure, and 2) controlling the agents for positioning and collision avoidance to establish that formation. 
When dealing with agents numbering in the thousands and millions for large structures, the complexity of the above fundamental problems and the difficulty of implementing such systems in the real-world increases exponentially. 
Agents suffer from heavy on-board computation leading to slow response times \cite{qiao2016consensus}, and implementation of large networks suffer from communication delays \cite{he2009robust}. 



Addressing the challenges listed by Mora \textit{et al.} towards structure formation by a large-scale multi-agent system, we present the following design constraints:
\begin{itemize}\label{requirements}
    \item A robust shape representation: The shape to be formed must not dictate the exact location of specific agents within the shape, 
    \item Achievable by simple, low cost, low overhead bearing agents following simple rules, and
    \item Achievable by local signaling or communication using small networks.
\end{itemize}

A number of works in literature has successfully addressed the shape representation requirement, but most fall short on the rest.
In this paper, a robust structure representation and a fully distributed structure formation methodology is proposed requiring simple individual agents with limited sensing and local communication for coordination, satisfying all the required constraints listed above. 


\section{Related Works}\label{relWork}
Over the years, a substantial amount of work has been proposed on multi-agent self-organization; a summary of representative works is presented in this section.
Oh \textit{et al.} categorized multi-agent shape formation as a position, displacement, or distance-based control problem in \cite{oh2015survey}. 
Formation control strategies based on relative positions of agents (distances and directions to neighbors) to maintain a shape have been presented in \cite{belta2004abstraction,gayle2009multi,anderson2008rigid,lee2014virtual}.
Prior work in the area has mainly focused on positioning agents in a pre-specified shape and evaluating the accuracy of the shape achieved \cite{antonelli2008entrapment}. 

Potential field-based approaches with global parameters to guide agents to goal locations over collision-free trajectories have been proposed in \cite{belta2004abstraction,gayle2009multi}. 
With the assumption of a multi-agent system network being jointly connected, collective motion patterns were obtained without any global beacon or guidance in \cite{zhang2016collaborative}.
Artistic pattern formation with visually appealing trajectories was proposed in \cite{alonso2011multi}.
Representative literature works on consensus based cooperative control towards shape formation include \cite{zhang2016collaborative,wang2017global,dong2015time,ding2017distributed,dong2017time}. 

Distributed coordination and control methods have been shown to achieve better scalability over centralized approaches. 
A distributed gradual pattern formation algorithm based on the Turing diffusion-driven instability theory was proposed in \cite{ikemoto2005gradual}.
Mactoobian \textit{et al.} showed that the overall computational workload in a multi-agent system can be probabilistically reduced by a distributed clustering approach \cite{macktoobian2017optimal}. 
Unnecessary information exchange between agents can also be minimized by event-triggered mechanisms in distributed cooperative control \cite{xu2016clustered,xu2018finite}. 
Other notable approaches toward improved scalability include finite time formation control strategies to address time constraint situations for large agent groups in \cite{ou2014finite,du2013finite}, and fixed time consensus control strategies studied in \cite{zuo2015nonsingular}. 
Predictor based control methods and consensus controllers have also been proposed in \cite{qiao2016consensus,wang2015consensus,zuo2017robust,wang2018predictor} to deal with input delays in large homogeneous/heterogeneous agent networks. 

The most successful implementation of large-scale multi-agent shape formation to date following local interactions have been achieved by the Kilobot platform \cite{rubenstein2012kilobot}; a programmable self-assembly of complex two-dimensional shapes with a thousand Kilobot swarm was successfully demonstrated in \cite{rubenstein2014programmable}. 
The strategy for forming 3D structures has not yet been fully realized. Unmanned aerial vehicles (UAVs) hold the benefit of superior maneuverability in 3D space, which aids in the formation of 3D structures. Previous work has mainly focused on the control of multi-UAV formations \cite{do2021formation}, but there has been little emphasis on generalizing 3D structure formation considering scalability.


The proposed method in this paper is in direct contrast with centralized controller-based multi-agent formation control, large-scale bidding and consensus-based approaches, and search-based inefficient methods where agents search for a place to settle in forming the required shape.
This proposed method in this paper addresses the issue of increasing computational requirements with an increasing number of agents by limiting coordination between agents to a local phenomenon using local sensing and communication, and allowing the settled agents to guide free-moving agents towards the formation frontier. This approach of guiding settling agents to a location and localized bidding through local interactions promote a highly scalable system in the context of large-scale implementation. 

\section{Preliminaries}\label{prelim}
\subsection{Structure Representation}
An arbitrary geometry made up of unit nodes is represented by a connected graph $\mathscr{G}=(V,E)$, where $V$ represents the set of $N$ nodes and $E$ represents the set of edges connecting neighboring nodes to form the structure without any self-connectivity following \cite{qureshi2007graph}. We represent node connections from node $D_k \in V$ to node $D_e \in V$, $\forall \{k,e\} \in E$ as $_k^{e}\{r,\theta,\psi\}$, defined as the distance, elevation and azimuth respectively of $D_e$ from $D_k$ relative to the parent node of $D_k$. The structure of the arbitrary shape can therefore be represented by a modified adjacency matrix $X$ with entries, 
\begin{align}\label{eq:xArg}
&X(m,n) =\begin{cases}
_m^{n}\{r,\theta,\psi\}      &\text{$\{m,n\} \in E$}\\
0                   &\text{else}
\end{cases}
\end{align}
where $m\in \{1,..,N\}$ and $n\in \{1,..,N\}$ denote row and column numbers respectively. 



\section{Methodology}\label{sol}

Assuming all agents in the system are aware of $X$, the structure formation process is started by placing an agent as the assigned initial beacon node of the structure with $X\{m,n\}\neq 0$ at the starting location of the to-be-formed structure. 
All other agents in the system can be randomly placed in the environment. 

Each agent is assumed to have a spherical detection and communication region of radius $r_d$ and $r_c$ respectively. At time $t$, $R_i$, $i \in \{1,..,N\}$ can either be staying in formation ($i \in A$), having detected the formation and moving towards it or moving along the structure surface ($i \in B$) or on random walk ($i \in C$). Here, $A$ is defined as the set of agent indices staying in formation, $B$ the set of agent indices moving towards the formation or along the structure surface, and $C$ the set of agent indices on random walk.
\subsection{Finite State Machine}
\begin{figure}[t]
	\centering
	\includegraphics[width=0.95\linewidth]{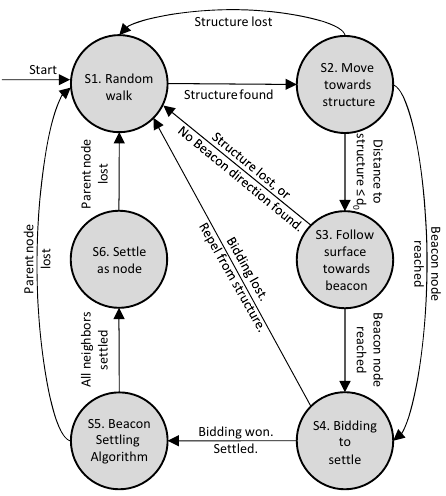}
  	\caption{Finite state machine controller with agent initial state S1.}
    \vspace{-6mm}
  	\label{fig:fsm_shapeForm}
\end{figure}
The shape formation process following the initial agent setup can be implemented as a finite state machine as shown in Fig. \ref{fig:fsm_shapeForm}. A description of the operation of this state machine is as follows.

\begin{figure}[t]     
	\centering
	\includegraphics[width=\linewidth]{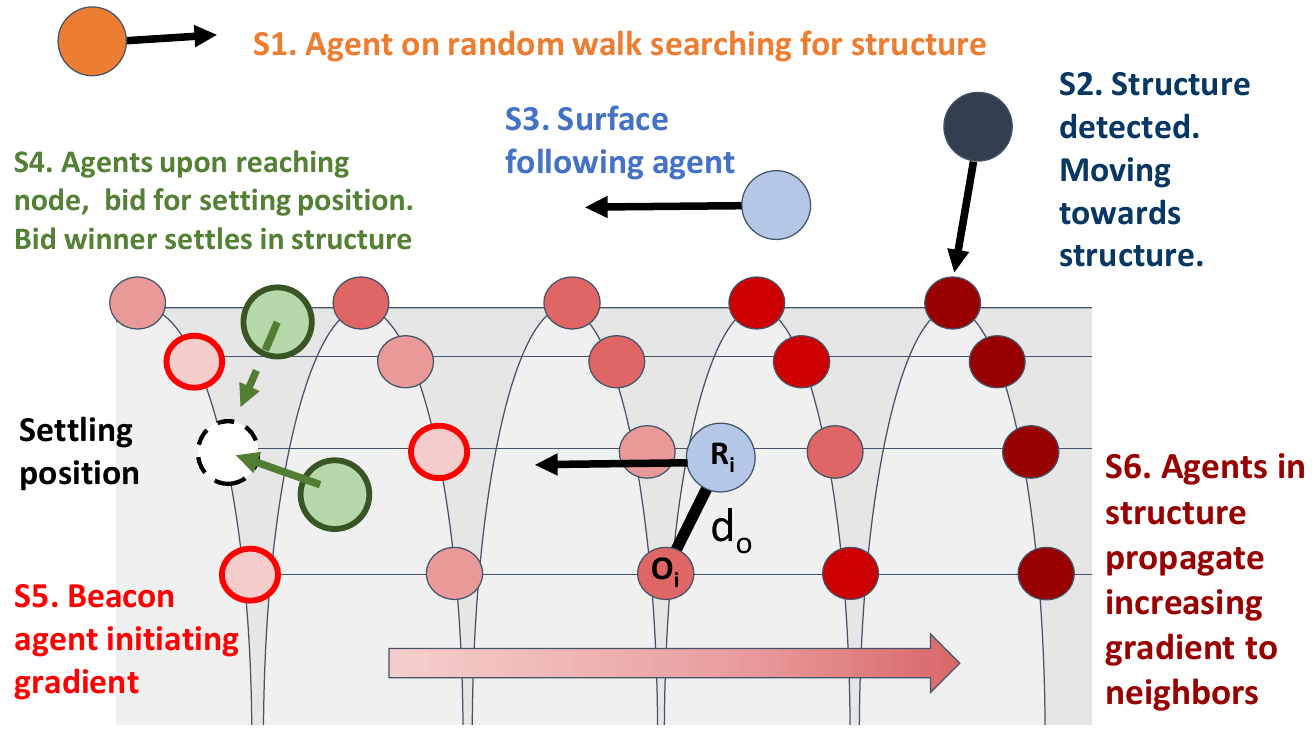}
  	\caption{Conceptual illustration showing the formation of structure in progress, with free-moving agent motion, surface gradient generation by beacons, agent surface-following motion, and node-neighbor interaction using the finite state machine controller.}
    \vspace{-8mm}
  	\label{fig:shape_build}
\end{figure}

\begin{itemize}
    \item \textbf{State S1: Search for shape structure.} $R_i, i\in C$ performs random walk in the environment looking for settled agents $R_j, j\in A$. Transitions:
    \begin{itemize}
        \item $S1 \longrightarrow S2$: if $R_j, j\in A$ found.
    \end{itemize}
    \item \textbf{State S2: Move towards structure.} $C \gets C\setminus\{i\}$, $B\gets B\cup \{i\}$. The nearest agent $R_j, j\in A$ from $R_i$ is set as the parent agent $O_i$ of $R_i$ with position denoted as $o_i$. $R_i, i\in B$ is attracted towards $O_i$. Transitions:
    \begin{itemize}
        \item $S2 \longrightarrow S4$: if $R_i, i\in B$ reaches surface following distance $||r_{i}o_i|| \leq d_0$ from parent $O_i$, where $d_0$ is the set surface following distance and the surface gradient value of $O_i$ is $b_{O_i} = 0$, suggesting $O_i$ is a beacon.
        \item $S2 \longrightarrow S3$: if $R_i, i\in B$ reaches surface following distance $||r_{i}o_i|| \leq d_0$ from parent $O_i$, where $d_0$ is the set surface following distance. 
        \item $S2 \longrightarrow S1$: if $O_i$ lost for time $T_l$.
    \end{itemize}
    \item \textbf{State S3: Follow surface towards beacon.} Agent $R_i,i\in B$ receives surface gradient values from all agents in structure within $r_d$, $R_j, j \in A$. It moves along the surface towards the direction of decreasing gradient from $O_i$ towards the nearest beacon maintaining distance $d_0$ from the structure surface. Transitions:
    \begin{itemize}
        \item $S3 \longrightarrow S4$: if the surface gradient value of $O_i$ is $b_{O_i} = 0$, suggesting $O_i$ is a beacon.
        \item $S3 \longrightarrow S1$: if $O_i$ lost for time $T_l$.
    \end{itemize}
    \item \textbf{State S4: Bidding to settle.} Agent $R_i,i\in B$ receives its possible node number $k$ from its beacon agent $O_i$ serving as node $p$ and communicates its bidding value $\epsilon_i$ to all other agents within $r_d$, $R_j \in B$ bidding for node $k$. Transitions:
    \begin{itemize}
        \item $S4 \longrightarrow S5$: if $\epsilon_i>\epsilon_j,\, \forall R_j \in B$ within $r_d$ of $R_i$ bidding for node $k$, communicates to $O_i$ that node $k$ has been settled.
        \item $S4 \longrightarrow S1$: if $\epsilon_i<=\epsilon_j,\, \forall R_j \in B$ within $r_d$ of $R_i$ bidding for node $k$, or $O_i$ lost for time $T_l$.
    \end{itemize}
    \item \textbf{State S5: Settle neighbors.} $B \gets B\setminus\{i\}$, $A\gets A\cup \{i\}$. $R_i$ moves to location $_p^{k}\{r,\theta,\psi\}$ relative to the position of $O_i$ and executes Algorithm \ref{alg:sn} to act as beacon for its neighboring agents to settle. Once all neighbors are settled, transitions:
    \begin{itemize}
        \item $S5 \longrightarrow S6$: all neighbors settled.
        \item $S5 \longrightarrow S1$: if $O_i$ lost for time $T_l$.
    \end{itemize}
    \item \textbf{State S6: Settle as node.} $R_i, i\in A$ settles in formation. Transitions:
    \begin{itemize}
        \item $S6 \longrightarrow S1$: if $O_i$ lost for time $T_l$.
    \end{itemize}
\end{itemize}
Figure \ref{fig:shape_build} illustrates the proposed shape formation strategy with the agent state and the surface gradient initiated by a beacon agent.

\subsection{Surface Gradient Propagation}\label{sec:surfaceProp}
Agents in S3 move along the surface of the structure towards the formation frontier following a surface gradient generated by beacon agents. 
Beacon agents initiate a gradient that is incrementally propagated throughout the structure by already settled agents.
The process has been previously presented in \cite{mamei2004experiments} inspired by morphogen gradients in biological sciences. 
A formal presentation of the process for agents in S3 with the required constraints is presented below.

We denote the gradient value of agent $R_i, i \in A$ as $b_{R_i}$. 
A beacon agent broadcasts its agent ID and an initial gradient value of $0$ within its limited spherical communication range with radius $r_c$. Having received the beacon agent gradient value of zero, all immediate neighbors re-broadcast their agent ID and an incremented gradient value of $1$. For multiple propagating gradients received, the minimum gradient value of all neighbors less than or equal to the current broadcast gradient value is incremented and broadcast as the new gradient value in the next time step. The process continues to propagate the incrementing gradient by subsequent neighbors. The continuous gradient propagation process where each agent broadcasts its gradient value in the next time step assessing the gradient values of all its immediate neighbors, can be formulated as,
\begin{equation}
    b_{R_i} = \text{min}\,b_{R_j} + 1, \quad j \in A | b_{R_j} < b_{R_i}^{prev}, r_{ij} \leq d_0
\end{equation}
where $b_{R_i}^{prev}$ is the gradient value broadcast by $R_i$ in the previous time step, and $r_{ij}$ is the Euclidean distance between agents $i,j \in \{A\}$. 
A color illustration of the surface propagated gradient from a beacon agent is shown in Fig. \ref{fig:shape_build}. 

\subsection{Beacon Settling Algorithm}\label{sec:beaconSettle}
\begin{algorithm}[t]
\caption{Beacon Settling Algorithm}\label{alg:sn}
\begin{algorithmic}[1]
\Procedure{}{$X, k, p$}
\State Communicate to $O_i$ of successful settlement as node $k$
\State Set beacon gradient value $b=0$
\For{$e=1 \to N$}
\If{$X(k,e) \neq 0$}
\While{Neighbor node $e$ settlement confirmation not received}
\State Broadcast $\{i,k,e,b\}$
\If{Neighbor node $e$ settled}
\State Break
\EndIf
\EndWhile
\EndIf
\EndFor
\EndProcedure
\end{algorithmic}
\end{algorithm}
Upon moving to S5, agent $R_i$ is aware of its parent's and its own node number on the shape structure mapping matrix $X$ denoted as $p$ and $k$ respectively. 
$R_i$ sequentially settles each of its neighbors following row $k$ of $X$ having non-zero entries and ignoring column $p$. 

The neighbor settling process is summarized in Algorithm \ref{alg:sn}. 
$R_i$ sets itself as the beacon ($b=0$) to propagate its own gradient over the forming structure surface. 
For every non-zero entry in row $k$ of the matrix $X$, $R_i$ broadcasts the message $\{i,k,e,b\}$ where $i$ is its own agent index, $k$ and $e$ are its self and the neighbor's node ID on the shape structure mapping matrix respectively, and $b=0$ is its initiating gradient value that surface following agents may read to detect it has a beacon. 
Once an agent settles as structure node ID $e$ and communicates that it has settled, $R_i$ moves on to settle its next neighbor following the shape structure mapping matrix $X$.

\subsection{Bidding for Node Position}\label{sec:bidding}
Agent $R_i$ in S3 having detected $O_i$ as a beacon receives the broadcasted beacon node and neighbor node numbers $p$ and $k$ respectively from the beacon. 
At any given time, more than one agent may reach a beacon. 
Each agent trying to occupy node $k$ of the structure bids for the position based on the distance travelled so far and its current distance to the location of node $k$. The bidding value may be calculated as, 
\begin{equation}
    \epsilon_i = \kappa_1\sum_{\tau=t_0+1}^t \,|r_i(\tau)-r_i(\tau-1)| + \kappa_2 r_{ik}    
\end{equation}
where $\kappa_1>0$ and $\kappa_2>0$ are scalar constants and $r_{ik}$ is the Euclidean distance between the agent location $r_i$ and location of the node $k$ denoted as $r_k$, determined from the shape structure mapping matrix $X$ and relative position of the beacon node $p$. 
All bidding agents within $r_d$ of one another broadcast and receive each other's bids. 
Agent $R_i$ individually evaluates its own bidding value against the rest to determine if the bid is won or lost. 
The agent with the winning bid moves in to occupy node $k$ of the structure by setting motion control constant $\eta=1$, while the rest return to state $S1$ after an initial repulsion with $\eta=-1$. Implementation details of $\eta$ are included in Section \ref{sec:motionCtrl}. 

\subsection{Agent Interaction and Motion Control}\label{sec:motionCtrl}
For modeling simplicity, we assume point mass dynamics for all agents without any maneuvering constraints. Agents in sets $A$ and $B$ maintain inter-agent distances using artificial potential $U_{ij}$, $i,j \in \{A,B\}$ previously established in \cite{leonard2001virtual} with an additional attraction term written as, 
\begin{align}\label{eq:Ui}
\underset{\forall i,j \in \{A,B\}}{U_{ij}} &= \begin{cases}
\alpha(\frac{1}{2}(r_{ij}-d_0)^2+\ln(r_{ij})+\frac{d_0}{r_{ij}}) &\text{$0<r_{ij}<d_1$}\\
\alpha(\frac{1}{2}(r_{ij}-d_1)^2+\ln(r_{ij})+\frac{d_0}{d_1}) &\text{$r_{ij}\geq d_1$}
\end{cases}
\end{align}
where $r_{ij}$ is the Euclidean distance between agents $i,j \in \{A,B\}$, $\alpha$ is a scalar control gain; $d_0$ and $d_1$ are scalar constants such that $d_0<d_1 \leq r_{d}$. Without dissipation, the structure equilibrium is locally stable in the sense of Lyapunov as shown in \cite{leonard2001virtual}.
We define $d_0$ as,
\begin{align}\label{eq:d1}
d_{0} &= \begin{cases}
d_{AA} &\text{$\forall i,j \in A$}\\
d_{AB} &\text{$\forall i \in A, j \in B$}
\end{cases}
\end{align}
such that $d_{AA} \leq d_{AB}$, where $d_{AA}$ and $d_{AB}$ are scalar constant parameters defining the inter-agent distances between agents in set $A$, and inter-agent distances in sets $A$ and $B$.
The inequality $d_{AA} \leq d_{AB}$ is defined to ensure that agents settled in structure maintain a compact lattice, and surface following agents maintain a larger safe distance from the structure for safety and maneuverability.

For State S4, agents are attracted to or repelled from a given relative node location based on bidding formulated as,
\begin{align}\label{eq:Uk}
U_{ik} &= \begin{cases}
\frac{\eta}{2}\beta r_{ik}^2 & \text{$R_i$ in S4}\\
0 &\text{else}
\end{cases}
\end{align}
where $r_{ik}$ is the Euclidean distance between $R_i$ and the relative location of the bidding node $k$ in the structure, $\beta$ is a scalar constant and $\eta \in \{-1,1\}$ for attraction or repulsion of $R_i$ from node location $k$ depending on the bidding outcome. 
The corresponding control input to maintain the desired distance between agents in sets A and B and in states S2, S4, S5 and S6 is defined as,
\begin{align}\label{eq:ui}
\underset{R_i \,\text{in}\, \{S2, S4, S5, S6\}}{u_i} &= \begin{cases}
-\sum_{j\neq i}\triangledown_{r_{ij}}U_{ij} +\triangledown_{r_{ik}}U_{ik}&\text{$0<r_{ij}<d_1$}\\
0 &\text{$r_{ij}\geq d_1$.}
\end{cases}
\end{align}

Agents currently in state S3 of the finite state machine and part of $B$, move along the structure surface in the direction of detected surface gradient following the control law, 
\begin{equation}\label{eq:Us3}
    \underset{R_i\,\text{in}\, S3}{u_{i}} = \frac{1}{2} \gamma \sum_{\forall j \in A | r_{ij} \leq r_d,\,b_{R_j}-b_{O_i}<0} |b_{R_j}-b_{O_i}|\, r_{io}\,\widehat{\overrightarrow{O_iR_j}}
\end{equation}
where $\gamma$ is a scalar constant, $r_{io}$ is the Euclidean distance between agents $R_i$ and $O_i$, and $\widehat{\overrightarrow{O_iR_j}}$ is the unit vector from agent $O_i$ to agent $R_j$ settled in the structure.

For agents in set C or state S1, random walk is achieved by setting a constant velocity $v_r$ at safe and bounded random elevation and azimuth orientations $\theta_r$ and $\psi_r$. 
Collision avoidance is implemented by agents by generating a new heading when another agent is detected within $r_d$.
We note here that the random walk implementation has been intentionally simplified in this paper for convenience as it is not the focus of the current work. 
Improved search methods by random walk with collision avoidance as presented in \cite{pang2019swarm} can be implemented in $S1$ for better performance.
\section{Stagnation Point Avoidance}\label{sec:stag}
Artificial potential-based motion control is susceptible to stagnation points, where the summation of all affecting potentials become zero, resulting in no net movement of the agent. 
However, agents in $A$ are exempt from this problem. We note the following strategies by agents in $B$ to escape possible stagnation points. 

The control input formulation in Eq. \ref{eq:ui} relies on the potentials defined in Eqs. \ref{eq:Ui} and \ref{eq:Uk}, where the gradient summation of these terms could result in a zero-control input.
Eq. \ref{eq:Us3} sums the gradient difference of neighboring agents from the closest node which may result in a net zero control input if the closest node $O$ is equidistant from two beacon nodes on opposite sides. 
In unique failure cases where all neighbors of the closest node agent within the detection range of a surface following agent have failed, the agent suffers from a net zero input.
In all such cases, agents in $B$ suffering from a net zero control input for a set time threshold returns to state $S1$ to generate a new path that allows it to escape the stagnation point. 
For multiple agents in such state clustered together on the structure surface, this process allows agents to move away from the structure to spread out and try again.
Agents follow the finite state machine to reacquire the structure and continue the structure formation process. 
The surface following process is in general robust to agent failures; surface following agents continue to move as long as at least one neighbor around its closest node broadcasts a gradient.
Therefore, the motion of surface following agents is concluded to be always in the direction of the nearest beacon even in the presence of possible stagnation points that may arise within the forming structure of the given shape. 

\section{Validation}\label{sec:expt_shape}
\begin{figure*}[ht!]
\centering
    \begin{subfigure}[t]{\textwidth}
    \centering
       \includegraphics[width=0.98\linewidth]{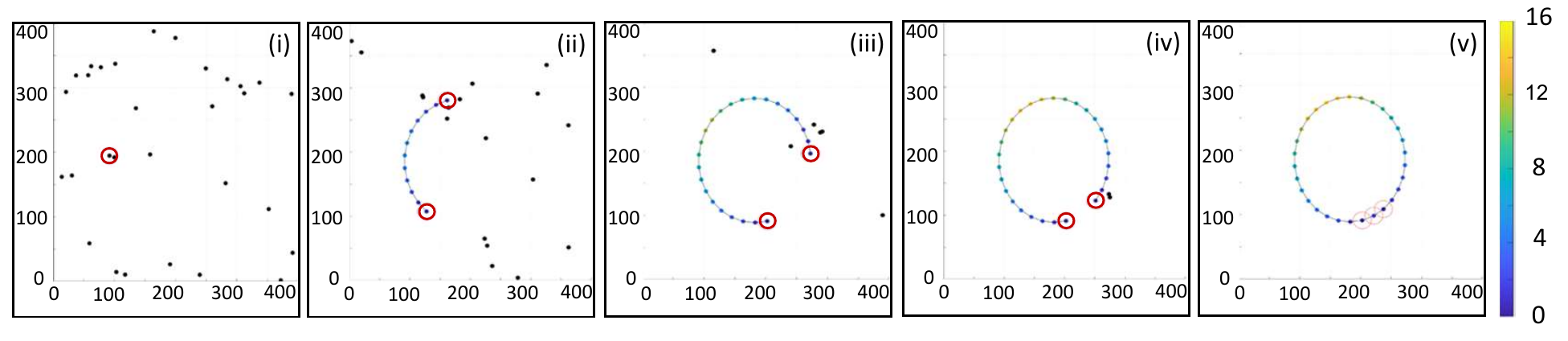}
       \vspace{-10pt}
        \caption{Circle shape formation with $N=30$ agents}
        \label{fig:circle_tl}
    \end{subfigure}
    \begin{subfigure}[t]{\textwidth}
    \centering
    \includegraphics[width=0.98\linewidth]{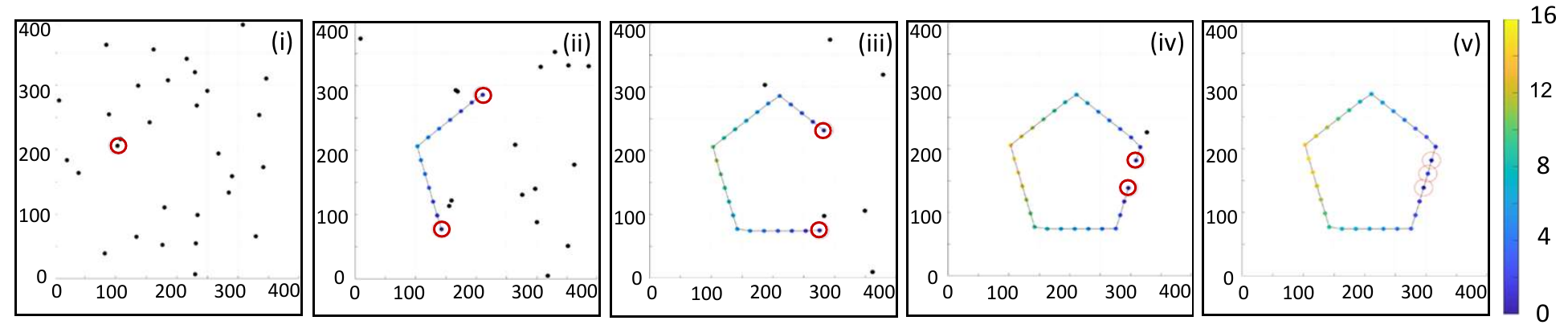}
         \vspace{-8pt}
        \caption{Pentagon shape formation with $N=30$ agents}
        \label{fig:pentagon_tl}
    \end{subfigure}
    \caption{Simulation set A: Planar shape formation time lapse shown in (i) - (v) with visual demonstration of the propagated surface gradient from the beacon agents. Beacon agents are marked in red. The proposed method successfully constructed shapes with convex continuously differentiable and discontinuous boundaries.}
    \vspace{-4mm}
    \label{fig:2d_tl}
\end{figure*}
We consider simple shapes such as a circle (with a continuously differentiable boundary) and a pentagon (with a discontinuous boundary) as the base shapes for the validation of the proposed structure formation strategy. 
The validation section is organized into two simulation sets. 
Simulation set A demonstrates planar shape formation, while simulation set B presents the 3D implementation of the structure formation process by extending the circle to a cylinder and the pentagon to a pentagonal prism. Scalability analysis is also included following the simulation results. 

For all simulation cases, all agents are randomly placed in a 3D space with dimensions of $400\,\times\,400\,\times\,400$, with an initial height of $10$ in the $z$ direction. 
The agents are free to move in any direction, with the simulation parameters set as follows: $r_d=40$, $d_1=40$, $d_{AA}=30$, $d_{AB}=30$, $\alpha=0.1$, $\beta=0.6$, and $\gamma = 0.3$, The random walk bound parameters are set as follows: $v_r=0.8$, $\theta_r=0$, and $\psi_r=\frac{\pi}{2}$. 
Agents are confined to remain within set bounds of the environment, by forcing them to turn back when crossing the area boundary. 
The formation process is initiated by placing the first node of the shape structure at the initializing location, which serves as the first beacon. 
A video of the simulations is available for reference at \url{https://youtu.be/o9aZbrkMAGA}.

\subsection{Simulation Set A: Circle and Pentagon}
A time-lapse of the planar circle and pentagon shape formation process for $N=30$ agents is presented in Fig. \ref{fig:2d_tl}. 
At initial time $t=0$, the initiating node is placed that acts as the first beacon following Algorithm \ref{alg:sn} attracting neighboring agents on random walk to settle and start the shape formation process. 
As more and more agents settle, the circle and pentagon shapes are observed to form over time from two ends until the last agent is attracted to its place, connecting the two ends to form the closed shape. 

Agents acting as beacon are highlighted with red circles. 
During the formation process, agents detecting the forming shape are attracted to the nearest agent already in the formation; boundary-following agents follow the gradient generated by the beacon agents along the length of the formed structure. 
The color scheme of the agents currently in the structure illustrates their broadcasted gradient values; beacon agents, highlighted in red circles, have a gradient value of zero, while the propagated gradient values reach as high as $15$ and $16$ on the far ends of the forming structure towards the end of the simulation for the circle and pentagon cases, respectively. 
The pentagon formation required traversing tight corners due to the discontinuous boundary of the shape. However, the agents were successfully able to follow the boundary and settle as part of the shape.


\subsection{Simulation Set B: Cylinder and Pentagonal Prism}
\begin{figure*}[ht]
\centering
    \begin{subfigure}[b]{\textwidth}
    \centering
       \includegraphics[width=0.95\linewidth]{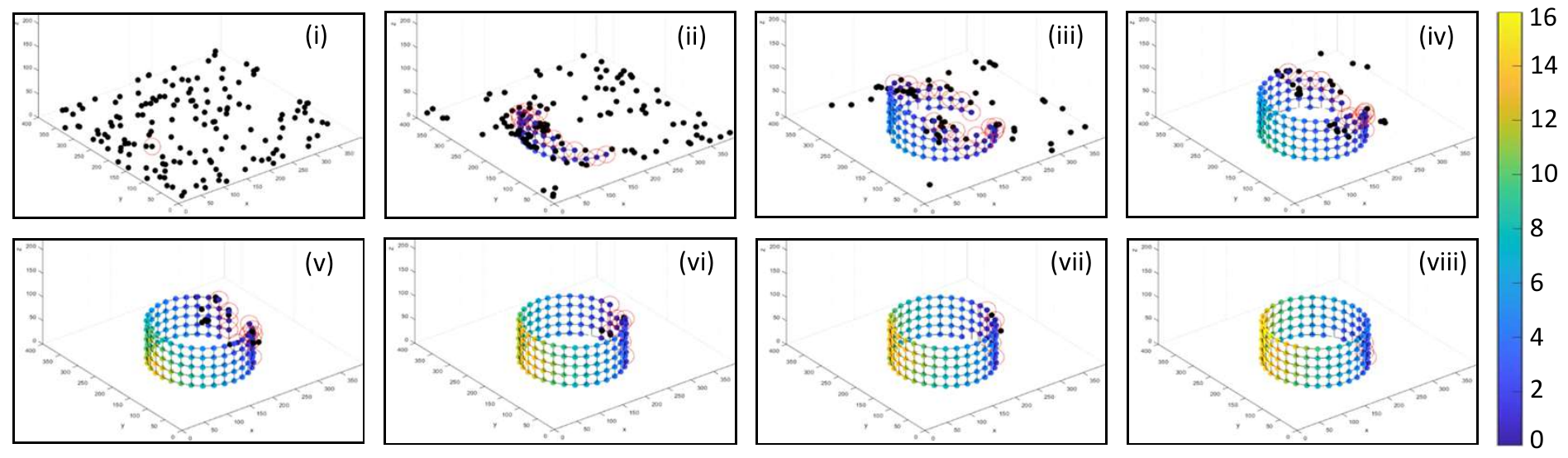}
        \vspace{-5pt}
        \caption{Cylindrical structure formation with $N=150$ agents}
        \label{fig:cylinder_tl}
    \end{subfigure}
    \begin{subfigure}[b]{\textwidth}
    \centering
        \includegraphics[width=0.95\linewidth]{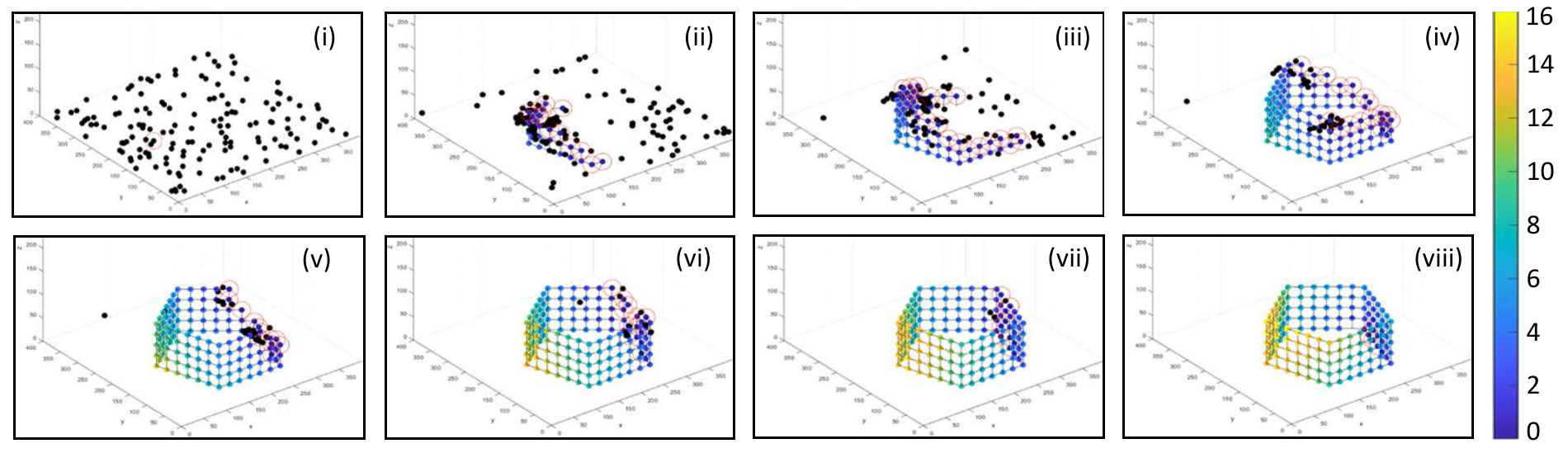}
        \vspace{-5pt}
        \caption{Pentagonal prism structure formation with $N=150$ agents}
        \label{fig:pentagonal_prism_tl}
    \end{subfigure}
    \caption{Simulation set B: 3D structure formation time lapse shown in (i) - (viii) with visual demonstration of the propagated surface gradient from the beacon agents. Beacon agents are marked in red. The proposed method successfully constructed structures with convex continuously differentiable and discontinuous surfaces.}
    \label{fig:3d_tl}
\end{figure*}
\begin{figure*}[h]
\centering
    \begin{subfigure}[b]{0.48\textwidth}
    \centering
       \includegraphics[width=0.95\linewidth]{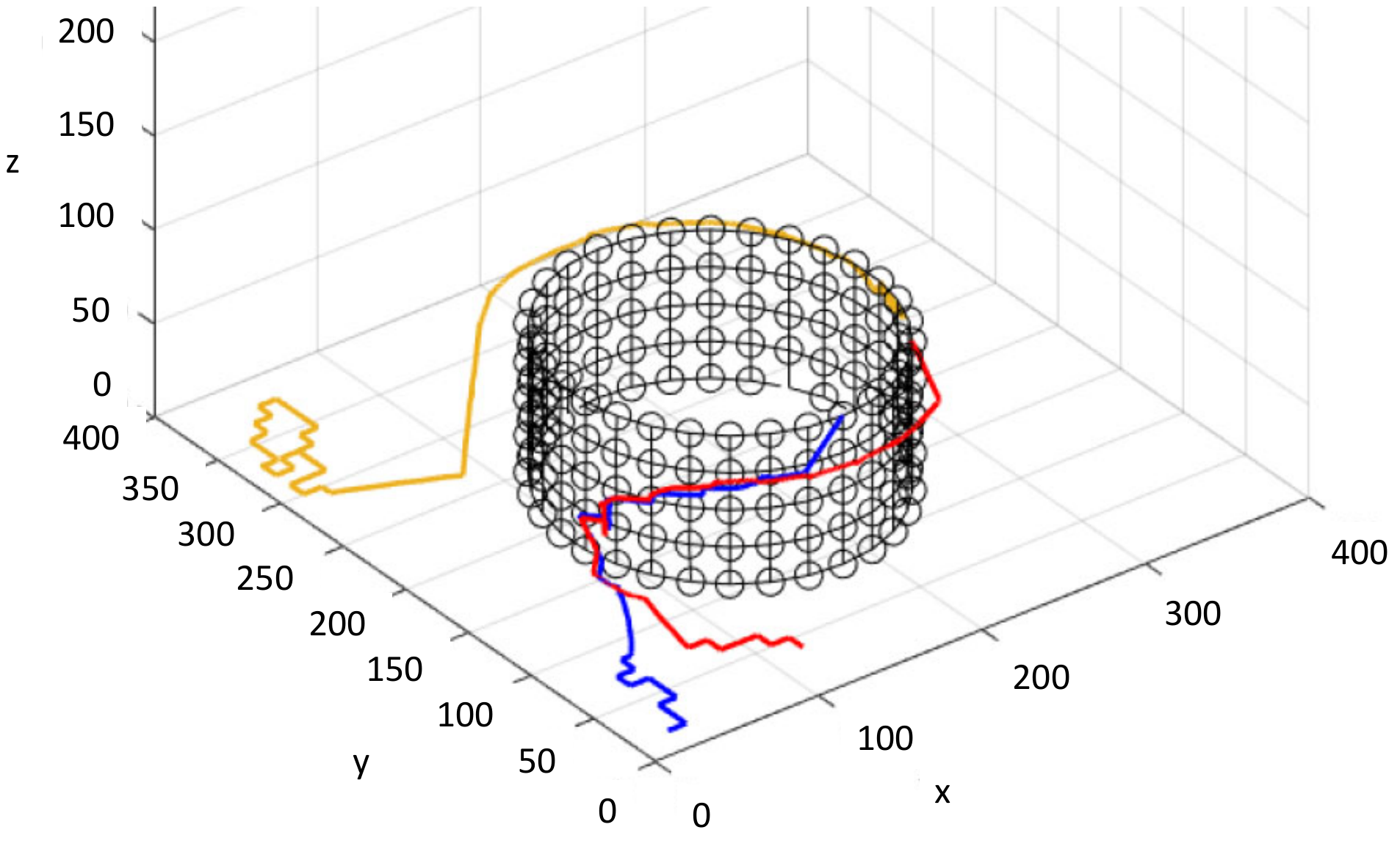}
        \caption{Cylinder formation}
        \label{fig:cylinder_track_iso}
    \end{subfigure}
    \begin{subfigure}[b]{0.48\textwidth}
    \centering
        \includegraphics[width=0.95\linewidth]{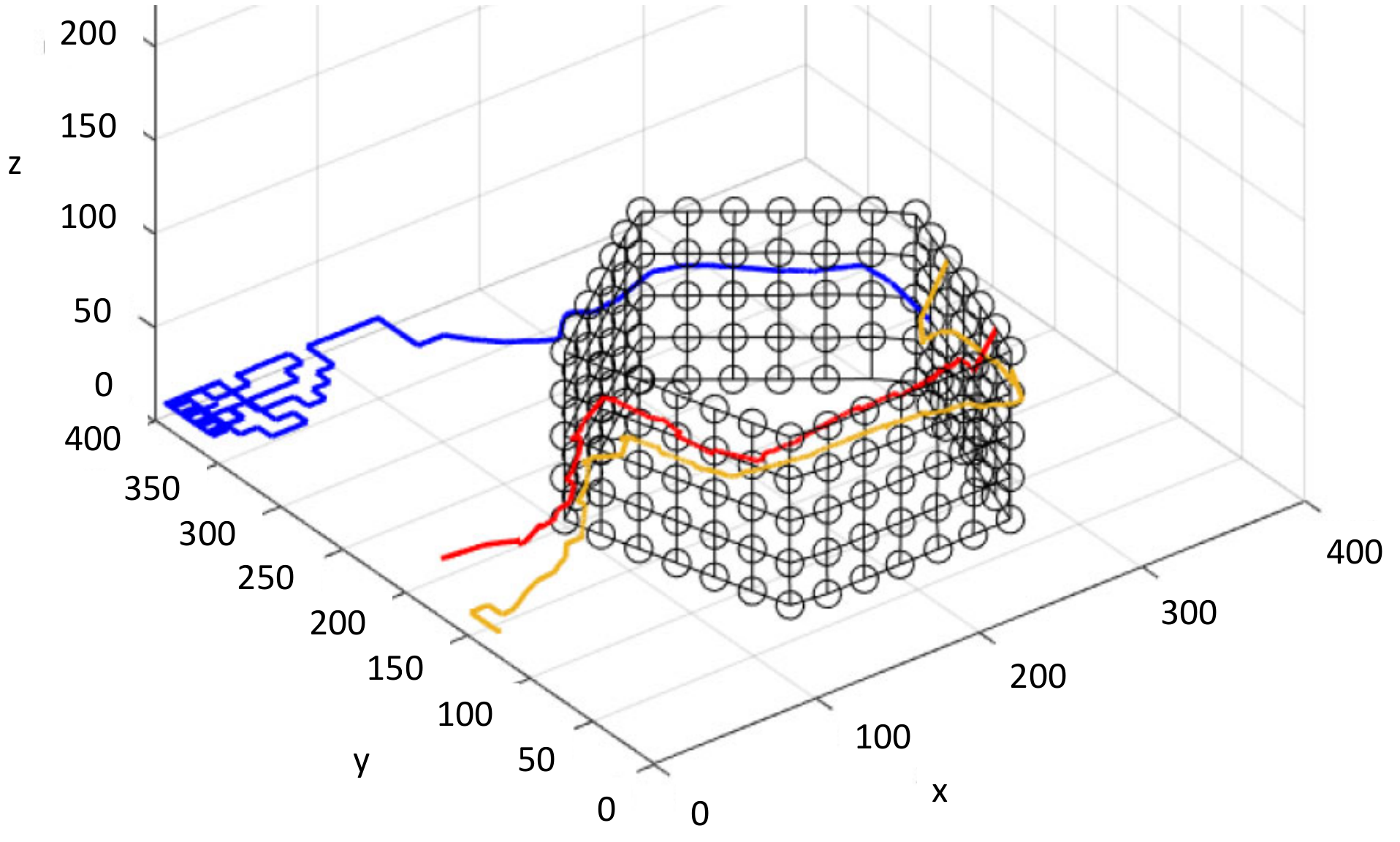}
        \caption{Pentagonal prism formation}
        \label{fig:pentagonPrism_track_iso}
    \end{subfigure}
    \caption{Paths taken by $3$ randomly picked agents in the cylinder and pentagonal prism formation process. Agents start with random walk and upon detection of the structure move towards it. Agents follow the time varying surface gradient to their settling locations successfully correcting their heading over time.}
    \vspace{-4mm}
    \label{fig:agent_3D_track}
\end{figure*}
A time-lapse of the 3D structure formation simulation for $N=150$ agents creating a cylinder and a pentagonal prism having a circular base and a pentagonal base with $30$ agents are shown in Fig. \ref{fig:3d_tl}. 
With the initiating agent placed as the first beacon for each case, neighboring agents on random walk were attracted to start the shape's structure formation process, similar to the planar simulation cases. 
After subsequent bidding rounds, agents were observed to settle at neighboring nodes, and the corresponding structures were observed to take shape over time. 
The propagated gradient along the surface of the forming structure is visualized by the color scheme. The beacon agents are seen in blue with a gradient value of zero, while the agents in structure on opposite ends reached a gradient value as high as $16$ over time shown in yellow.
All agents successfully followed the proposed distributed finite state machine controller to form the respective 3D structures.

The path taken by $3$ randomly selected agents in the formation process is presented in Fig. \ref{fig:agent_3D_track}. 
The agents started with random walk and upon detecting the structure are observed to move towards it.
The agents follow the surface gradient to reach a position to settle.
Since the beacon agents change over time in the structure formation process, the observed surface gradient by free moving agents is time-varying in nature; therefore, agents update their path accordingly. 
This phenomenon is observed in the cylinder formation process in Fig. \ref{fig:cylinder_track_iso}, where the red and blue agents, upon reaching the structure, initially move in one direction from level $1$ of the structure to level $2$ and then change their heading, moving past level $2$. Similar observations can be made for the pentagonal prism formation in Fig. \ref{fig:pentagonPrism_track_iso}.

\subsection{Scalability Analysis}\label{sec:scalability}
\begin{figure}[t]     
	\centering
	\includegraphics[width=0.95\linewidth]{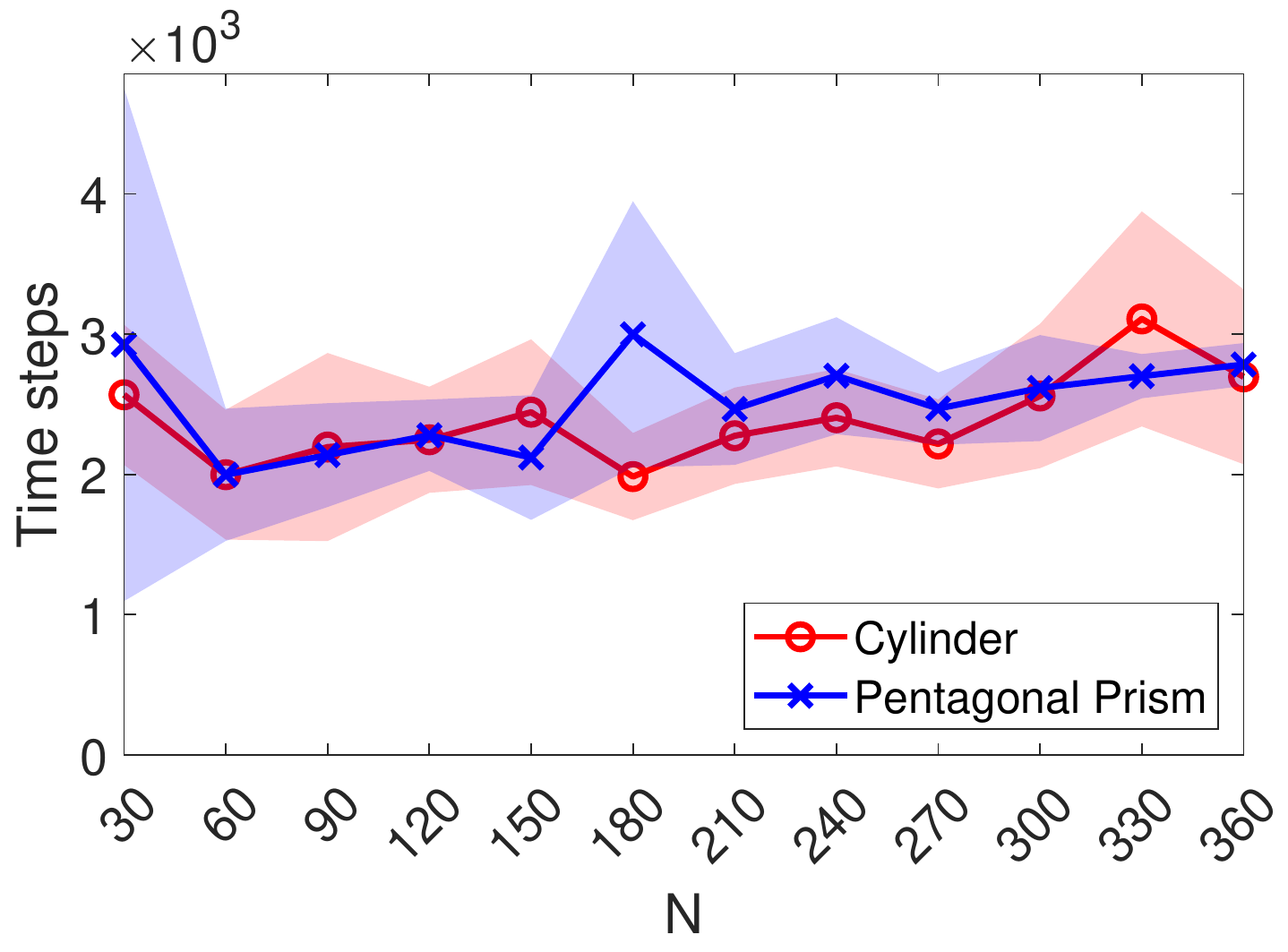}
  	\caption{Plot of the mean number and standard deviation of simulation time steps for the formation of cylinder and pentagonal prism structures with increasing $N$.}
    \vspace{-4mm}
  	\label{fig:scalability}
\end{figure}
We demonstrate the scalability of the proposed structure formation method by repeating simulation set B, which involves the formation of a cylinder and pentagonal prism structure, for an increasing number of $N$ up to $360$ at increments of $30$. 
Each case of $N$ for the cylinder and pentagonal prism formation was repeated $5$ times, and the mean and standard deviation of all trials are plotted in Fig. \ref{fig:scalability} with increasing $N$.

Under the constraints of the closed simulation environment, the required number of time steps consistently increased with increasing $N$ in both the cylinder and pentagonal prism formation processes.
Once the structure formation process is initiated, larger structures tend to have a larger number of active beacon agents compared to small structures at any given time aiding the formation on multiple fronts. Therefore, the difference in required time steps between the $N=30$ to $N=360$ structures is observed to be fairly small for the presented simulation set.
The $N=30$ case for both the structures is presented as outliers requiring a larger mean number of time steps. For smaller $N$ and a smaller structure, the number of beacon agents settling neighbors was the smallest in the simulation set, and agents remained longer in state S1 trying to find the forming structure, contributing to the higher number of required formation time steps.
Although further investigation is required on more complex shapes, based on our current findings, we conclude that the proposed method is highly scalable in the multi-agent domain for the formation of simple building block structures.

\subsection{Discussion}
The surface following process is specifically designed for continuous surfaces without any sharp edges in the structure. 
However, the agents were observed to be robust to traversing \textit{soft} edges of shapes such as the pentagons and pentagonal prisms successfully in all trials.
Similar observations were made for polygons with more than five sides.
However, the system struggled to form structures for polygons having fewer  than five sides, such as rectangles and rectangular prisms. 
Sharp edges on these shapes created stagnation points that forced agents to repeatedly switch to state $S1$ and try again.
Further investigation is required on the proposed method in determining the maximum allowed exterior angle of surface edges. 
The proposed methodology could also be improved by formulating the motion of surface following agents to independently move around sharp edges to the other side, or updating the structure representation formulation to accommodate structures requiring sharp edges or corners to be rounded with a higher agent density, such that a continuous surface could be achieved for the surface following process.    

The proposed system yielded fairly efficient paths taken by each agent during their motion along the structure surface following the surface gradient to reach the beacon agents. 
We stress here the importance of using this gradient following method as opposed to fixed motion patterns such as raster scanning along aligned agents, helical motion of surface following agents in the vertical direction or random motion to eventually reach a beacon agent looking for a neighbor to settle. While fixed motion methods may require minimal computation and less or no communication in determining motion direction, the paths taken by the agents are likely to be highly inefficient in comparison without a sense of how and where the structure is currently forming. 

Since the proposed system utilizes a completely distributed process where each agent determines its own actions based on received communication and observations of immediate neighbors, the system is concluded to be robust against agent motion failures. 
Since any position in the structure may be taken by any agent, the shape formation process will continue and complete with the remaining agents as long as their paths are not blocked by disabled agents. 
The simulation was repeated for special cases where clusters of agents in the structure broadcasting their propagated gradient values were disabled to simulate their failure cases. 
Surface following agents were observed to simply move around these failed agent clusters without any broadcasted gradients; the failed agents were treated as agents not in structure, and therefore, no surface was present to follow. Special cases of stagnation points were also validated, where a surface following agent was unable to find a motion direction when its nearest node was broadcasting, but all its neighbors were simulated to fail. 
Following the stagnation point avoidance criteria in Section \ref{sec:stag}, the surface following agent simply returned to random walk to escape and find another surface agent still broadcasting. 
However, it should be noted that the proposed system is unable to account for beacon agent failures to settle a neighbor that is not accounted for by any other beacons in the vicinity.       

The proposed research is focused on developing a generic multi-agent coordination method for 3D structure formation. 
Direct comparison of our work with existing implementations such as the Kilobot shape formation \cite{rubenstein2014programmable} and Termes system \cite{werfel2011distributed} is beyond the scope of this work at this stage without further development considering dynamic and kinematic constraints of existing platforms; therefore, we leave that as future work. 

\section{Conclusion}\label{conclude}
In this paper, we present a distributed self-organizing strategy for large-scale multi-agent systems to form prescribed shapes and structures using local communication and sensing with immediate neighbors. 
The simulation results presented demonstrate the effectiveness of the proposed strategy in forming simple planar and 3D shapes. 
The proposed method is scalable to form the structure of a given shape assuming an adequate number of agents are available in the system.
The simulation experiments assumed perfect localization and sensing by the modeled agents. We leave the investigation of the effects of noise on the proposed strategy as future work.\\



\addtolength{\textheight}{-5cm}   




\bibliographystyle{IEEEtran}
\bibliography{references}

\begin{thebibliography}{10}
\providecommand{\url}[1]{#1}
\csname url@samestyle\endcsname
\providecommand{\newblock}{\relax}
\providecommand{\bibinfo}[2]{#2}
\providecommand{\BIBentrySTDinterwordspacing}{\spaceskip=0pt\relax}
\providecommand{\BIBentryALTinterwordstretchfactor}{4}
\providecommand{\BIBentryALTinterwordspacing}{\spaceskip=\fontdimen2\font plus
\BIBentryALTinterwordstretchfactor\fontdimen3\font minus
  \fontdimen4\font\relax}
\providecommand{\BIBforeignlanguage}[2]{{%
\expandafter\ifx\csname l@#1\endcsname\relax
\typeout{** WARNING: IEEEtran.bst: No hyphenation pattern has been}%
\typeout{** loaded for the language `#1'. Using the pattern for}%
\typeout{** the default language instead.}%
\else
\language=\csname l@#1\endcsname
\fi
#2}}
\providecommand{\BIBdecl}{\relax}
\BIBdecl

\bibitem{alonso2011multi}
J.~Alonso-Mora, A.~Breitenmoser, M.~Rufli, R.~Siegwart, and P.~Beardsley,
  ``Multi-robot system for artistic pattern formation,'' in \emph{2011 IEEE
  International Conference on Robotics and Automation}.\hskip 1em plus 0.5em
  minus 0.4em\relax IEEE, 2011, pp. 4512--4517.

\bibitem{qiao2016consensus}
W.~Qiao and R.~Sipahi, ``Consensus control under communication delay in a
  three-robot system: Design and experiments,'' \emph{IEEE Transactions on
  Control Systems Technology}, vol.~24, no.~2, pp. 687--694, 2016.

\bibitem{he2009robust}
X.~He, Z.~Wang, and D.~Zhou, ``Robust fault detection for networked systems
  with communication delay and data missing,'' \emph{Automatica}, vol.~45,
  no.~11, pp. 2634--2639, 2009.

\bibitem{oh2015survey}
K.-K. Oh, M.-C. Park, and H.-S. Ahn, ``A survey of multi-agent formation
  control,'' \emph{Automatica}, vol.~53, pp. 424--440, 2015.

\bibitem{belta2004abstraction}
C.~Belta and V.~Kumar, ``Abstraction and control for groups of robots,''
  \emph{IEEE Transactions on robotics}, vol.~20, no.~5, pp. 865--875, 2004.

\bibitem{gayle2009multi}
R.~Gayle, W.~Moss, M.~C. Lin, and D.~Manocha, ``Multi-robot coordination using
  generalized social potential fields,'' in \emph{2009 IEEE International
  Conference on Robotics and Automation}.\hskip 1em plus 0.5em minus
  0.4em\relax IEEE, 2009, pp. 106--113.

\bibitem{anderson2008rigid}
B.~D. Anderson, C.~Yu, B.~Fidan, and J.~M. Hendrickx, ``Rigid graph control
  architectures for autonomous formations,'' \emph{IEEE Control Systems
  Magazine}, vol.~28, no.~6, pp. 48--63, 2008.

\bibitem{lee2014virtual}
Y.-H. Lee, S.-G. Kim, T.-Y. Kuc, J.-K. Park, S.-H. Ji, Y.-S. Moon, and Y.-J.
  Cho, ``Virtual target tracking of mobile robot and its application to
  formation control,'' \emph{International Journal of Control, Automation and
  Systems}, vol.~12, no.~2, pp. 390--398, 2014.

\bibitem{antonelli2008entrapment}
G.~Antonelli, F.~Arrichiello, and S.~Chiaverini, ``The entrapment/escorting
  mission,'' \emph{IEEE Robotics \& Automation Magazine}, vol.~15, no.~1, pp.
  22--29, 2008.

\bibitem{zhang2016collaborative}
H.-T. Zhang, Z.~Chen, and M.-C. Fan, ``Collaborative control of multivehicle
  systems in diverse motion patterns,'' \emph{IEEE Transactions on Control
  Systems Technology}, vol.~24, no.~4, pp. 1488--1494, 2016.

\bibitem{wang2017global}
B.~Wang, J.~Wang, B.~Zhang, and X.~Li, ``Global cooperative control framework
  for multiagent systems subject to actuator saturation with industrial
  applications,'' \emph{IEEE Transactions on Systems, Man, and Cybernetics:
  Systems}, vol.~47, no.~7, pp. 1270--1283, 2017.

\bibitem{dong2015time}
X.~Dong, B.~Yu, Z.~Shi, and Y.~Zhong, ``Time-varying formation control for
  unmanned aerial vehicles: Theories and applications,'' \emph{IEEE
  Transactions on Control Systems Technology}, vol.~23, no.~1, pp. 340--348,
  2015.

\bibitem{ding2017distributed}
Z.~Ding, ``Distributed adaptive consensus output regulation of
  network-connected heterogeneous unknown linear systems on directed graphs,''
  \emph{IEEE Transactions on Automatic Control}, vol.~62, no.~9, pp.
  4683--4690, 2017.

\bibitem{dong2017time}
X.~Dong, Y.~Zhou, Z.~Ren, and Y.~Zhong, ``Time-varying formation tracking for
  second-order multi-agent systems subjected to switching topologies with
  application to quadrotor formation flying,'' \emph{IEEE Transactions on
  Industrial Electronics}, vol.~64, no.~6, pp. 5014--5024, 2017.

\bibitem{ikemoto2005gradual}
Y.~Ikemoto, Y.~Hasegawa, T.~Fukuda, and K.~Matsuda, ``Gradual spatial pattern
  formation of homogeneous robot group,'' \emph{Information Sciences}, vol.
  171, no.~4, pp. 431--445, 2005.

\bibitem{macktoobian2017optimal}
M.~Macktoobian and M.~Aliyari~Sh, ``Optimal distributed interconnectivity of
  multi-robot systems by spatially-constrained clustering,'' \emph{Adaptive
  Behavior}, vol.~25, no.~2, pp. 96--113, 2017.

\bibitem{xu2016clustered}
W.~Xu and D.~W. Ho, ``Clustered event-triggered consensus analysis: An
  impulsive framework,'' \emph{IEEE Transactions on Industrial Electronics},
  vol.~63, no.~11, pp. 7133--7143, 2016.

\bibitem{xu2018finite}
W.~Xu, Z.~Wang, and D.~W. Ho, ``Finite-horizon $h_\infty$ consensus for
  multiagent systems with redundant channels via an observer-type
  event-triggered scheme,'' \emph{IEEE transactions on Cybernetics}, vol.~48,
  no.~5, pp. 1567--1576, 2018.

\bibitem{ou2014finite}
M.~Ou, H.~Du, and S.~Li, ``Finite-time formation control of multiple
  nonholonomic mobile robots,'' \emph{International Journal of Robust and
  Nonlinear Control}, vol.~24, no.~1, pp. 140--165, 2014.

\bibitem{du2013finite}
H.~Du, S.~Li, and X.~Lin, ``Finite-time formation control of multiagent systems
  via dynamic output feedback,'' \emph{International Journal of Robust and
  Nonlinear Control}, vol.~23, no.~14, pp. 1609--1628, 2013.

\bibitem{zuo2015nonsingular}
Z.~Zuo, ``Nonsingular fixed-time consensus tracking for second-order
  multi-agent networks,'' \emph{Automatica}, vol.~54, pp. 305--309, 2015.

\bibitem{wang2015consensus}
C.~Wang, Z.~Zuo, Z.~Lin, and Z.~Ding, ``Consensus control of a class of
  lipschitz nonlinear systems with input delay,'' \emph{IEEE Transactions on
  Circuits and Systems I: Regular Papers}, vol.~62, no.~11, pp. 2730--2738,
  2015.

\bibitem{zuo2017robust}
Z.~Zuo, C.~Wang, and Z.~Ding, ``Robust consensus control of uncertain
  multi-agent systems with input delay: a model reduction method,''
  \emph{International Journal of Robust and Nonlinear Control}, vol.~27,
  no.~11, pp. 1874--1894, 2017.

\bibitem{wang2018predictor}
C.~Wang, Z.~Zuo, Z.~Qi, and Z.~Ding, ``Predictor-based extended-state-observer
  design for consensus of mass with delays and disturbances,'' \emph{IEEE
  transactions on cybernetics}, no.~99, pp. 1--11, 2018.

\bibitem{rubenstein2012kilobot}
M.~Rubenstein, C.~Ahler, and R.~Nagpal, ``Kilobot: A low cost scalable robot
  system for collective behaviors,'' in \emph{2012 IEEE International
  Conference on Robotics and Automation}.\hskip 1em plus 0.5em minus
  0.4em\relax IEEE, 2012.

\bibitem{rubenstein2014programmable}
M.~Rubenstein, A.~Cornejo, and R.~Nagpal, ``Programmable self-assembly in a
  thousand-robot swarm,'' \emph{Science}, vol. 345, no. 6198, pp. 795--799,
  2014.

\bibitem{do2021formation}
H.~T. Do, H.~T. Hua, M.~T. Nguyen, C.~V. Nguyen, H.~T. Nguyen, H.~T. Nguyen,
  and N.~T. Nguyen, ``Formation control algorithms for multiple-uavs: a
  comprehensive survey,'' \emph{EAI Endorsed Transactions on Industrial
  Networks and Intelligent Systems}, vol.~8, no.~27, pp. e3--e3, 2021.

\bibitem{qureshi2007graph}
R.~J. Qureshi, J.-Y. Ramel, and H.~Cardot, ``Graph based shapes representation
  and recognition,'' in \emph{International Workshop on Graph-Based
  Representations in Pattern Recognition}.\hskip 1em plus 0.5em minus
  0.4em\relax Springer, 2007, pp. 49--60.

\bibitem{mamei2004experiments}
M.~Mamei, M.~Vasirani, and F.~Zambonelli, ``Experiments of morphogenesis in
  swarms of simple mobile robots,'' \emph{Applied Artificial Intelligence},
  vol.~18, no. 9-10, pp. 903--919, 2004.

\bibitem{leonard2001virtual}
N.~E. Leonard and E.~Fiorelli, ``Virtual leaders, artificial potentials and
  coordinated control of groups,'' in \emph{Decision and Control, 2001.
  Proceedings of the 40th IEEE Conference on}, vol.~3.\hskip 1em plus 0.5em
  minus 0.4em\relax IEEE, 2001, pp. 2968--2973.

\bibitem{pang2019swarm}
B.~Pang, Y.~Song, C.~Zhang, H.~Wang, and R.~Yang, ``A swarm robotic exploration
  strategy based on an improved random walk method,'' \emph{Journal of
  Robotics}, vol. 2019, pp. 1--9, 2019.

\bibitem{werfel2011distributed}
J.~K. Werfel, K.~Petersen, and R.~Nagpal, ``Distributed multi-robot algorithms
  for the termes 3d collective construction system,'' in \emph{Proceedings of
  Robotics: Science and Systems}.\hskip 1em plus 0.5em minus 0.4em\relax
  Institute of Electrical and Electronics Engineers, 2011.

\end{thebibliography}

\end{document}